\documentclass[aps,prc,a4paper,twocolumn]{revtex4}
\usepackage{graphicx}
\begin{document}

\title{\bf Semi-classical and anharmonic quantum models of nuclear wobbling motion}
\author{Makito Oi}
\email[e-mail: ]{ m.oi@surrey.ac.uk}
\affiliation{Department of Physics, University of Surrey,
Guildford, GU2 7XH, United Kingdom}

\date{14 December, 2005 (Ver.1.1)}

\begin{abstract}
A semi-classical model for  wobbling motion is presented 
as an extension to the Bohr-Mottelson model of wobbling motion.
Using the resultant wobbling potential,
a quantum mechanical equation is derived for  anharmonic wobbling motion.
We then attempt to explain the anharmonicity 
observed in the excited bands of two wobbling phonons 
in the $A\simeq 160$ region.
\end{abstract}
\maketitle
\addvspace{7mm}

Significant progress has been made in a last couple of years
with regard to the understanding of nuclear wobbling motion.
This phenomenon was first predicted theoretically by Bohr and Mottelson 
thirty years ago \cite{BM75}, but
it was not observed until experimental techniques advance such
as 4$\pi$ gamma-ray detectors. 
Finally in 2001, the first experimental report was published
 on the evidence for wobbling motion in $^{163}$Lu \cite{OHJ01}.
Subsequently, possible wobbling excitations were reported
 also in $^{165,167}$Lu \cite{SHH03,Am03}.
These wobbling bands (of one-phonon excitation) were analyzed first 
by the particle-rotor model (PRM) \cite{OHJ01,Ha02}
and then by the random phase approximation (RPA) \cite{MSM02}.

Evidence for two-phonon bands has also been reported in
$^{163,165}$Lu \cite{DRJ02,Am03}. Despite much experimental evidence
implying that the bands consist of two-phonon excitations,
a serious discrepancy with the original Bohr-Mottelson model
was observed in the experimental energy spectrum:
it shows strong anharmonicity.
Currently, neither PRM nor RPA are successful to explain this
anharmonicity, but Matsuzaki and Ohtsubo proposed an interesting idea 
with the tilted-axis cranking model (TAC) to account for this
anharmonicity \cite{MO04}. 
They calculated the energy surface with respect to the tilt angles
of the total angular momentum vector, and found that the curvature
of the energy surface around the origin becomes flatter 
as angular momentum increases.
At (and beyond) some critical angular momentum, the local minimum at the origin
disappears and new minima emerge as a manifestation of tilted rotation. 
They suggested that this ``phase transition'' 
from  principal-axis rotation (PAR)
to tilted-axis rotation (TAR) causes the anharmonicity,
although they did not explicitly demonstrate a mechanism 
for the excited bands to acquire the anharmonic character. 

The aim of this paper is to present an analytical model to explain
the anharmonicity observed in the nuclear wobbling motion.
Two steps are required to derive the anharmonic wobbling model proposed
in this work.
First, we consider a semi-classical treatment of the Bohr-Mottelson model
to derive the wobbling potential energy. Then, the re-quantisation
is made using the potential, so as to obtain the quantum mechanical
equation for the anharmonic wobbling motion.

In the Bohr-Mottelson model, the Hamiltonian reads
\begin{equation}
  \hat{H}=\sum_{i=1}^3\frac{\hat{I}^2_i}{2{\cal J}_i}.
\end{equation}
This is the Hamiltonian of a triaxial quantum rotor
without intrinsic structure. Classically, the dynamics
of this rotor can be described by the three Euler angles 
and three components of  the total angular momentum vector. 
The angular momentum operators $\hat{I}_i$ in the Hamiltonian
are those in the
body-fixed frame, so that their commutation relations are given as
\begin{math}
  [\hat{I}_i,\hat{I}_j]=-i\epsilon_{ijk}\hat{I}_k,
\end{math}
where $\epsilon_{ijk}$ denotes the Levi-Civita symbol.
The Hamiltonian commutes with the total angular momentum
$\hat{I}^2=\sum_{i=1}^3\hat{I}_i^2$, but each component $\hat{I}_i$
does not commute when the system possesses triaxiality (${J}_1 >{J}_2 >{J}_3$, 
for instance).
\begin{equation}
 [\hat{H},\hat{I}_i] = \frac{i}{2}
 \left(\frac{1}{\mathcal{J}_j}-\frac{1}{\mathcal{J}_k}\right)
 \left(2\hat{I}_j\hat{I}_k+i\hat{I}_i\right) \ne 0,
\end{equation}
with $(i,j,k)$ being cyclic.
In the Bohr-Mottelson model, the quantization axis is chosen
along the 1-axis. The above non-commutation relation implies
that the eigenvalue of $\hat{I}_1$ (denoted in this paper as $K$) 
is not a good quantum number. 
Therefore, it is convenient to introduce here the following expression 
for the wobbling state, which is generally described as
\begin{equation}
  |\text{wobble};I\rangle = \sum_{K=-I}^{I}C_K|IK\rangle.
  \label{wbl-state}
\end{equation}

In order to allow the classical analogy, the high-spin condition
is applied in the Bohr-Mottelson model. Classically, this condition
is expressed as  $I\simeq I_1 >> 1$. Considering the non-conservation of
the $K$ quantum number, this classical condition should be interpreted as 
\begin{equation}
  I\simeq \langle\hat{I}_1\rangle >> 1,
  \label{cls-cond}
\end{equation}
where the expectation value is taken with respect to the state
given in Eq.(\ref{wbl-state}). To satisfy this condition, 
we require now
\begin{equation}
  |C_{I}|^2 >> \sum_{K\ne I}|C_{K}|^2.
  \label{wbl-cnd}
\end{equation}
Due to the normalisation condition $\sum_{K}|C_K|^2=1$,
the above equation also implies $|C_{I}|^2\simeq 1$.

Now, the Hamiltonian is decomposed into two terms.
\begin{equation}
  \hat{H}=\frac{I(I+1)}{2{\cal J}_1} + \hat{H}_{\text{w}},
\end{equation}
where the Dirac constant is set to be unity ($\hbar=1$).
The first term is a c-number because $I$ is the quantum number
(the total angular momentum). The second term contains q-numbers,
\begin{equation}
  \hat{H}_{\text{w}}=\frac{1}{2}
  \left(\frac{1}{{\cal J}_2}-\frac{1}{{\cal J}_1}\right)
  \hat{I}_{2}^2
+
\frac{1}{2}
  \left(\frac{1}{{\cal J}_3}-\frac{1}{{\cal J}_1}\right)
  \hat{I}_{3}^2.
\end{equation}
In the Bohr-Mottelson model,
bosonic operators of creation and annihilation are introduced so as to
diagonalise $\hat{H}_{\text{w}}$. Two steps are necessary for this aim.
First, the creation and annihilation operators are respectively defined as
\begin{equation}
  a^{\dag}=\frac{\hat{I}_2+i\hat{I}_3}{\sqrt{2I}} \quad \text{and} \quad
  a=\left(a^{\dag}\right)^{\dag}.
\end{equation}
Second, to eliminate so-called ``dangerous terms''
 (such as $a^{\dag}a^{\dag}$ and $aa$),
the canonical transformation is performed to introduce a new representation
of the creation-annihilation operators $(c,c^{\dag})$.
\begin{equation}
  c^{\dag}=xa^{\dag}-ya, \quad \text{with} \quad x^2-y^2=1.
\end{equation}
Here $x$ and $y$ are written in terms of the moment of inertia (${\cal J}_i$)
and are determined so as to make the dangerous terms in the new representation
vanish. 
The commutation relation is invariant with respect
to the canonical transformation. 
The pair of operators $(c,c^{\dag})$ therefore follows 
the same commutation relation as the original. 
The resultant Hamiltonian can be written by only c-numbers.
The corresponding energy spectrum now reads
\begin{equation}
  E^{I}_n=\frac{I(I+1)}{2{\cal J}_1}+\omega_{\text{w}}\left(n+\frac{1}{2}\right).
\end{equation}
Here $n$ is an eigenvalue of the number operator $\hat{n}\equiv c^{\dag}c$
and the wobbling excitation energy is given as 
\begin{math}
  \omega_{\text{w}}=\left(I/{\cal J}_1\right)
\sqrt{{\left({\cal J}_1-{\cal J}_2\right)
\left({\cal J}_1-{\cal J}_3\right)}/\left({\cal J}_2{\cal J}_3\right)}.
\end{math}

It looks as if the Hamiltonian is fully quantised, but it is the
result of the approximation for the bosonic commutation relation,
\begin{equation}
  [a,a^{\dag}]= \hat{I}_1/I \simeq 1,
  \label{boson}
\end{equation}
which is based on the condition Eq.(\ref{cls-cond}).
The exact form of $\hat{H}_{\text{w}}$ should be, 
\begin{equation}
  \hat{H}_{\text{w}}=\omega_{\text{w}}
  \left(\hat{n}+\frac{\hat{I}_1}{2I}\right).
  \label{q-wbl-hml}
\end{equation}
Because we can easily prove that $\hat{n}$ and $\hat{I}_1$ do not
commute with each other, $n$ cannot be treated as a quantum number
without the approximation in Eq.(\ref{boson}).

This approximation is a little drastic from a quantum mechanical point of
view, not only because the operator $\hat{I}_1$ is simply replaced by
a c-number ($K$), but also because the c-number is a non-conserved 
$K$-quantum number in this model. However, this approximation can be
totally justified when we consider the expectation values of the
relevant quantities. That is, instead of Eqs.(\ref{boson}) and 
(\ref{q-wbl-hml}), we have
\begin{eqnarray}
  \langle[a,a^{\dag}]\rangle &=&\langle\hat{I}_1\rangle/I \simeq 1, \\
  \langle\hat{H}_{\text{w}}\rangle &=& \omega_{\text{w}}
  \left(\langle\hat{n}\rangle + \frac{\langle\hat{I}_1\rangle}{2I}\right) \\
  &\simeq &
  \omega_{\text{w}}
  \left(\langle\hat{n}\rangle + \frac{1}{2}\right),
\end{eqnarray}
where the expectation value is taken with respect to the wobbling state
given in Eq.(\ref{wbl-state}).
This ``semi-classical'' approximation cannot be compared with the experimental
energy spectrum at low-lying states directly 
because the number of wobbling phonons can now take
any real number which gives a continuous spectrum. In other words, 
the current model is set back to the classical theory due to the 
semi-classical approximation.

To re-quantise the system,
let us introduce a dynamical variable $\theta$ which is defined as
\begin{equation}
  \theta = \cos^{-1}\left(\frac{\langle\hat{I}_1\rangle}{I}\right).
\end{equation}
This dynamical variable physically means the wobbling angle of the
total angular momentum vector.
If the high-spin condition $I\simeq \langle\hat{I}_1\rangle >> 1$
is valid, $\theta$ is very small ($\theta << 1$),
which is the case considered in the original wobbling model.
The average value for the number operator is calculated as,
\begin{equation}
  \langle\hat{n}+\frac{1}{2}\rangle \simeq
  \left(\frac{1}{4}\kappa I+\frac{1}{2}\right)\left(1-\frac{\langle\hat{I}_1\rangle}{I}
  \right),
\end{equation}
where the term $\text{Re}\left(C_IC_{I-2}\right)$ is neglected due to
the condition Eq.(\ref{wbl-cnd}) and $\kappa$ is given as
\begin{equation}
  \kappa = \frac{1/{\cal J}_2+1/{\cal J}_3-2/{\cal J}_1}
         {\sqrt{\left(1/{\cal J}_2-1/{\cal J}_1\right)
         \left(1/{\cal J}_3-1/{\cal J}_1\right)}}.
\end{equation}
In this way, the mean value of the wobbling Hamiltonian
is now expressed as
\begin{eqnarray}
&&  \langle\hat{H}_{\text{w}}\rangle \\ \nonumber
  &&= \omega_{\text{w}}\left[\frac{\kappa I}{4}
    \left\{1-\frac{\langle\hat{I}_1\rangle}{I}\left(1-\frac{1}{I}\right)\right\}
    +\frac{1}{2}\left(1-\frac{\langle\hat{I}_1\rangle}{I}\right)\right],\\ \nonumber
  &&\simeq \omega_{\text{w}}\left(\frac{\kappa I}{4}+\frac{1}{2}\right)
    \left(1-\frac{\langle\hat{I}_1\rangle}{I}\right).
\end{eqnarray}
Because it is expected that the order of $\kappa$ is of $O(1)$,
we may justify writing $\langle\hat{H}_{\text{w}}\rangle$ as
\begin{eqnarray}
  \label{anh-pot}
  \langle\hat{H}_{\text{w}}\rangle &\simeq & 
  \frac{\kappa I}{4}\omega_{\text{w}}
    \left( 1-\cos\theta\right) \\ \nonumber
    &=& \frac{I^2}{8}\left(1/{\cal J}_2+1/{\cal J}_3-2/{\cal J}_1\right)
        \left(1-\cos\theta\right).
\end{eqnarray}

Let us consider the classical motion of $\theta$ by
treating this energy expectation value 
as the potential energy for the classical wobbling motion.
The associated Lagrangian is now introduced as
\begin{equation}
  L(\theta,\dot{\theta}) = \frac{1}{2}m\dot{\theta}^2 - 
  m\omega_{\text{sc}}^2\left(1-\cos\theta\right).
\end{equation}
The semi-classical oscillator frequency $\omega_{\text{sc}}$ is 
introduced here as
\begin{equation}
  m\omega_{\text{sc}}^2 = \frac{I^2}{8}\left(1/{\cal J}_2+1/{\cal J}_3-2/{\cal J}_1\right).
  \label{coeff}
\end{equation}
This interpretation is possible 
because under the high-spin condition given in Eq.(\ref{cls-cond}), 
that is, when the wobbling angle is small ($\theta << 1$), 
the potential is approximated that of
a simple harmonic oscillator, that is, 
$V(\theta)\simeq \frac{1}{2}m\omega_{\text{sc}}^2\theta^2$.

Therefore, quantisation for the wobbling motion 
is carried out through the simple one-dimensional Schr\"odinger equation.
\begin{equation}
 \left\{  -\frac{1}{2m}\frac{d^2}{d\theta^2} +
 m\omega_{\text{sc}}^2\left(1-\cos\theta\right)\right\}
 \Psi(\theta) = E_{\text{w}}\Psi(\theta).
 \label{anh-eq}
\end{equation}
This equation can be considered as an extended wobbling model
that can handle even the wobbling motion with large amplitude
(that is, $I >> \langle\hat{I}_1\rangle$). It should be noted
that we did not use the first part of 
the approximation Eq.(\ref{cls-cond}) in the derivation.
(The high-spin condition $I>>1$ is used, however.)

In this model, the mass parameter $m$ is the only free parameter.
However, using experimental data, 
we can estimate the order of $m$ in the following way. 
First, when $\theta <<1$, the energy spectrum
is given as $E_{\text{w}}(\nu)=\omega_{\text{sc}}\left(\nu+\frac{1}{2}\right)$.
The one-phonon excitation energy ($\Delta E^{\text{1-phonon}}$) 
observed in $^{163}$Lu is approximately 350 keV,
which should be equal to $\omega_{\text{sc}}$.
Using Eq.(\ref{coeff}), we have
\begin{equation}
  m=
\frac{I^2}{8}
\frac{\left(1/{\cal J}_2+1/{\cal J}_3-2/{\cal J}_1\right)}
{\left(\Delta E^{\text{1-phonon}}\right)^2}.
\label{mass-param}
\end{equation}

The anharmonic wobbling potential derived in our model,
that is, $V(\theta) \propto 1-\cos\theta$, is not  strong enough
to reproduce the anharmonicity observed in  experiment. 
According to experiment, the excitation energy from the first
phonon state to the second is reduced by nearly 50\% in comparison
to the excitation energy from the yrast to the first phonon state,
whereas our anharmonic wobbling spectrum gives, at most, 10\% reduction
(at $I=30\hbar$), 
from the result of our numerical calculation where the rigid body
moment of inertia is used for $^{163}$Lu with $(\beta,\gamma)=(0.4,-20^{\circ})$.
(The degree of triaxiality
is here measured by $\gamma$ \cite{RS80}. In this study,
we employ the Hill-Wheeler coordinates, which gives the opposite sign
convention to the so-called Lund convention.)
The mass parameter $m$ is positive for the rigid-body moment of
inertia with $\gamma \alt 35^{\circ}$, which applies to the present case.
Owing to this fact, we consider 
only the case that the mass parameter is positive ($m>0$),
in the following discussions.
According to the spin-dependence in Eq.(\ref{mass-param}), the mass parameter
becomes increased at higher spin.
At $I=30\hbar$ and $40\hbar$, 
$m$ is calculated as $m=6.7$ and 11.9 (MeV$^{-1}$),  respectively. 
Consequently,
the higher the total angular momentum, the less anharmonicity.
For instance, the reduction is less than 5\% at $I=40\hbar$.
In addition, with irrotational-flow moment of inertia , we found that 
the anharmonicity turns out to be much less substantial.
This is because the moment of inertia is proportional to $\beta^2$
in the case of irrotational flow, which gives rise to 
one order of magnitude smaller in comparison to the rigid body case.
As a result, $m$ tends to be one order of magnitude larger
in accordance with Eq.(\ref{mass-param}).

In the rest of this paper, we attempt to explain the discrepancy 
between the above model and observed spectrum, which shows very strong 
anharmonicity.
One possibility for the discrepancy could be the
evolution of nuclear structure originating from microscopic degrees of freedom,
which is neglected in the model. For example, such an effect can be
attributed to the Coriolis force in the rotating frame.
Not only can the stretching effect of the nuclear shape cause a
change of the moment of inertia, but also the quasi-particle excitations
due to rotational alignment cause a kind of ``phase transition''
from the BCS (or superfluid) phase to the normal fluid phase.
Therefore, in order to investigate further, we might need to
go to microscopic theories such as the 
3D-cranked Hartree-Fock-Bogoliubov approach and the
generator coordinate method \cite{OAH00},
which demand massive numerical efforts.

\begin{figure}[bth]
  \includegraphics[width=0.45\textwidth]{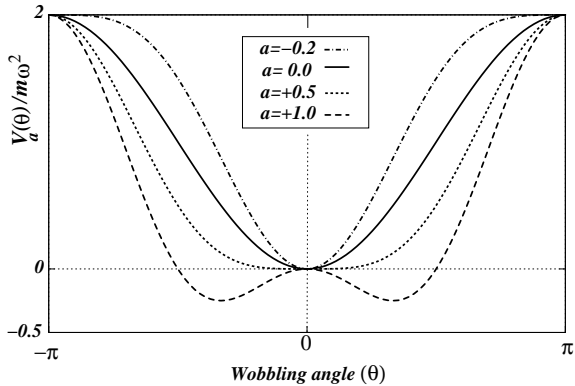}
  \caption{The wobbling potential energy with strong anharmonicity, 
    $V_a(\theta)$, for different control parameter, $a$.}
    \label{pot}
\end{figure}

The suggestion by Matsuzaki and Ohtsubo based on their
microscopic calculations is useful in considering possible effects
originating from intrinsic structure without performing
demanding numerical calculations.
In the following, we focus on a ``phase transition'' from PAR to TAR,
as mentioned at the beginning of this paper.
In order to account for very strong anharmonicity, 
let us consider the following potential, 
which is similar to the energy surface obtained by Matsuzaki and Ohtsubo.
\begin{equation}
 V_{a}(\theta) \propto 1-\cos\theta - a\left(1-\cos^2\theta\right).
\end{equation}
Fig.\ref{pot} shows the shape of the potential for difference choices of the
parameter, $a$.
When $\theta << 1$, the above potential reduces to
$V_a(\theta)\simeq (1-2a)\theta^2/2$, which implies the effective mass 
$m^*=(1-2a)m$. 
From the simple analytical analysis of the above potential,
one can tell that the critical point happens at $a=1/2$. (See Fig.\ref{pot})
Below the critical 
point ($-1/2< a<1/2$), the classical ground state,
that is, the minimum of $V(\theta)$, is found at $\theta=0$.
This solution corresponds to PAR. Whereas,
beyond the critical point ($a>1/2$), the classical solutions appear
at $\theta=\pm \cos^{-1}(1/2a)$, which correspond to TAR.
Namely, the parameter $a$ has a physical meaning as a control parameter
of a (classical) phase transition between TAR and PAR.
(The domain $a<-1/2$ is not considered in this study
because an unphysical minimum appears at $\theta=\pm\pi$.) 
\begin{figure}[tbh]
  \includegraphics[width=0.35\textwidth,angle=-90]{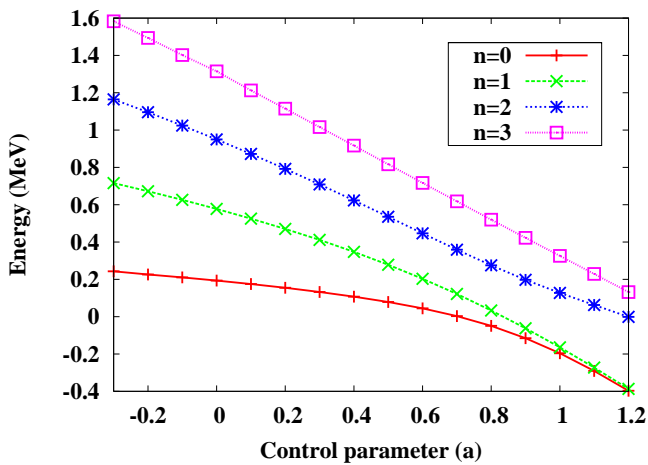}
  \caption{Wobbling spectrum as a function of the control parameter, $a$.
    The rigid-body moment of inertia is employed with 
  $(\beta,\gamma)=(0.4,-20^{\circ}).$ }
\end{figure}

The physical situation seen beyond the critical point ($a>1/2$)
is similar to the mechanism of 
spontaneous symmetry breaking in the linear sigma model, or $\phi^4$ theory
in quantum field theory \cite{Ry85}. 
Unlike  $\phi^4$ theory, however, the cause of the
``negative mass'' in our case can be attributed to more specific
physical effects, that is,  microscopic effects. 
It is widely known that nuclear moment of inertia matches neither
that of a rigid body, nor of an irrotational flow. It takes, in fact, the value
between these two limiting cases \cite{RS80}. The primary reason is 
the presence of the pairing correlation between constituent particles 
(nucleons). Because the mass parameter $m$ in our rotor model is determined 
by the moment of inertia in such a way presented in Eq.(\ref{mass-param}),
the deviation of the realistic value of the moment of inertia from the
rigid-body or irrotational-flow values can influence the magnitude of
the mass parameter. 
It is therefore natural that 
the microscopic effects, such as the pairing correlation,
 are taken into account through the renormalisation of the mass parameter,
 that is, $m^*$.
In other words,
the microscopic effects are supposed to be included in 
the present model effectively, through the control parameter, $a$.
When the effective mass parameter, $m^*$ 
changes from a positive value to negative,
one can suppose that the spontaneous symmetry breaking occurs
as an analogy with the quantum field theory.
A difference from the $\phi^4$ theory is that the broken symmetry in our
model is a discrete symmetry ($\theta\rightarrow -\theta$), so that
no Nambu-Goldstone mode \cite{Nam} is created 
for the restoration of the symmetry.
Instead,  quantum tunnelling plays the role of the symmetry restoration.

To examine how the strong anharmonicity is induced through the spontaneous
symmetry breaking, numerical calculations are carried out  for
the potential $V_a(\theta)$ by varying the control parameter, $a$.
The results are shown in Fig.2.
Below the critical point ($a<1/2$), the energy difference between the 
ground and the first excited state is of the order of $\hbar\omega_{\text{sc}}$,
which is about 0.35 MeV in the present case. The anharmonicity can be seen,
but they are very small (less than 10\%), 
as already discussed earlier in connection to the original wobbling potential,
Eq.(\ref{anh-pot}).
However, beyond the critical point, the energy difference becomes extremely 
small due to the tunnelling effect. As the ground state is bound deeper in the
potential, the energy splitting becomes smaller. Finally, around $a\simeq 1.1$,
the ground and the first excited states can be regarded as
almost degenerate, which implies an onset of the spontaneous symmetry breaking
in a quantum mechanical sense.
In other words,
 the quantised wobbling motion around $\theta=0^{\circ}$ (PAR) 
goes into a phase transition to TAR beyond the critical point ($a\simeq 1.1$).
Considering that the degeneracy is perfect, the lowest three levels
($n=1,2,$ and 3) give rise to a very strong anharmonicity.
In particular, at $a=1.1$, the ratio is calculated to be
$\Delta E_{2\rightarrow 1}/\Delta E_{3\rightarrow 2} =$ 335 (keV)/ 167 (keV) 
$\simeq 2$,
where $\Delta E_{i\rightarrow j}\equiv E_j-E_i$. This result agrees 
well with the experimentally observed anharmonicity.

These ideas are based on the consideration that
microscopic degrees of freedom bring further anharmonicity
in addition to the wobbling potential derived in Eq.(\ref{anh-pot}).
It is thus necessary to check the ideas and models presented here
by using microscopic approaches, which has to be done in the future.

In summary, a semi-classical wobbling model is presented about an extension
of the original wobbling model by Bohr and Mottelson to derive
the wobbling potential energy. This potential is anharmonic and 
described by a wobbling angle $\theta$, 
which is introduced as a dynamical variable to represent
the semi-classical wobbling motion. The re-quantisation is made with this
potential and the dynamical variable. The quantum equation
for the anharmonic wobbling motion is then derived.
However, this anharmonic potential cannot reproduce the experimental
spectrum. In order to explain the discrepancy, a further anharmonic model
taking into account a phase transition going into TAR from PAR
is proposed as a possible influences coming from the intrinsic microscopic
degrees of freedom and the associated quantum tunnelling effect.
With the proper choice of the control parameter, it is demonstrated 
qualitatively that such a strong anharmonicity seen in experiment can be
well reproduced as a consequence of the spontaneous symmetry breaking.
\begin{acknowledgements}
In constructing the strong anharmonic wobbling potential, $V_a(\theta)$,
the critical-point theory developed by Iachello in \cite{Ia00}
was very useful.
The author thanks Prof. H. Flocard and Prof. R. C.
Johnson for fruitful discussions.
Careful reading of the manuscript by Prof. P. M. Walker is acknowledged, too.
Financial support from an EPSRC advanced research fellowship GR/R75557/01
and an EPSRC grant EP/C520521/1 are also appreciated by the author.
\end{acknowledgements}

\end{document}